\documentclass[journal]{IEEEtran}
\usepackage{mathtools}
\usepackage{amsmath}
\usepackage{amssymb}
\usepackage{amsthm} 
 \usepackage{multirow}
\usepackage{stfloats}
 \usepackage{array}  
\usepackage{booktabs} 
 \usepackage{amssymb}
\usepackage{amsmath, cases}
\usepackage{graphicx}
\usepackage{epstopdf}
\usepackage{multicol}
\usepackage{algorithmic}
\usepackage{algorithm}
\usepackage{graphicx}
\usepackage{subfigure}
\usepackage{float}
\usepackage{comment}
\usepackage{verbatim}
\usepackage{cite}
\usepackage{color}
\usepackage{amssymb}
\usepackage{mathrsfs}
\usepackage{amsfonts,amssymb}
\usepackage{caption}
\usepackage{fancyhdr}
\usepackage{amsmath,amssymb}
\usepackage{amsfonts}
\usepackage{xfrac}
\usepackage{extarrows}
\usepackage[hidelinks]{hyperref}    

\usepackage{color}
\usepackage{pgfplots}           
\pgfplotsset{compat=1.16}
\usepackage{enumerate}
\usepackage{graphics}
\usepackage{subfigure}
\usepackage{cite}
\usepackage{mathrsfs}
\usepackage{caption}
\usepackage{stfloats}

\usepackage{tikz}
\usepackage{mathdots}
\usepackage{yhmath}
\usepackage{cancel}
\usepackage{color}
\usepackage{siunitx}
\usepackage{array}
\usepackage{multirow}
\usepackage{textcomp}
\usepackage{gensymb}
\usepackage{tabularx}
\usepackage{booktabs}
\usepackage{graphicx}
\usepackage{xcolor}

\newtheorem{theorem}{Theorem}

\newtheorem{proposition}{Proposition}

\begin{document}

\title{On the Ergodic Rate of Cognitive Radio Inspired Uplink Multiple Access}
\author{Xiao Yue,
Sotiris A. Tegos,~\IEEEmembership{Student Member,~IEEE,}
Panagiotis D. Diamantoulakis,~\IEEEmembership{Senior Member,~IEEE,}\\
Zheng Ma,~\IEEEmembership{Senior Member,~IEEE,} and George K. Karagiannidis,~\IEEEmembership{Fellow,~IEEE}
\thanks{This work was supported by the European Regional
Development Fund by the European Union and Greek national
funds through the Operational Program Competitiveness,
Entrepreneurship and Innovation, under the call: Special
Actions: Aquaculture - Industrial Materials - Open Innovation
in Culture (project code: T6YBP-00134).}
\thanks{Y. Xiao and Z. Ma are with the Provincial Key Lab of Information Coding and Transmission, Southwest Jiaotong University, Chengdu 610031, China (e-mails: alice\_xiaoyue @hotmail.com; zma@home.swjtu.edu.cn).}
\thanks{S. A. Tegos is with the Wireless Communications and Information Processing (WCIP) Group, Electrical \& Computer Engineering Department, Aristotle University of Thessaloniki, 54124, Thessaloniki, Greece (e-mails: tegosoti@auth.gr).}
\thanks{P. D. Diamantoulakis and G. K. Karagiannidis are with the Wireless Communications and Information Processing (WCIP) Group, Electrical and Computer Engineering Department, Aristotle University of Thessaloniki, 54124,  Thessaloniki, Greece, and also with the Provincial Key Lab of Information Coding and Transmission, Southwest Jiaotong University, Chengdu 610031, China (e-mails: padiaman@ieee.org; geokarag@auth.gr).} 
\vspace{-0.2in}}
\maketitle
%

\begin{abstract}  With the exponential increase of the number of devices in the communication ecosystem toward the upcoming sixth generation (6G) of wireless networks, more enabling technologies and potential wireless architectures are necessary to fulfill the networking requirements of high throughput, massive connectivity, ultra reliability, and heterogeneous quality of service (QoS).
In this work, we consider an uplink network consisting of a primary user (PU) and a secondary user (SU) and, by integrating the concept of cognitive radio and multiple access, two protocols based on rate-splitting multiple access and non-orthogonal multiple access with successive interference cancellation are investigated in terms of ergodic rate. 
The considered protocols aim to serve the SU in a resource block which is originally allocated solely for the PU without negatively affecting the  QoS  of  the  PU.
We extract the ergodic rate of the SU considering a specific QoS for the PU for the two protocols. In the numerical results, we validate the theoretical analysis and illustrate the superiority of the considered protocols over two benchmark schemes.
\end{abstract}



\vspace{-0.2in}
\section{Introduction}
\IEEEPARstart{W}{ith} the development of the Internet of Things (IoT) and the consequent integration of a huge amount of heterogeneous wireless devices, sharing the same orthogonal resources is necessary to achieve the goals of the next generation of wireless communication networks, including higher connectivity and data rates, as well as reduced delay and energy consumption  \cite{ZZQ19VTM, Liu2022}.
To efficiently explore enhanced networking functionalities in such complicated wireless communication scenarios and meet the heterogeneity of networking quality of service (QoS) requirements, enabling network slicing techniques, evolutionary multiple access mechanisms, and advanced cognitive radio (CR) technologies are required. 
The use of this concept as the enabler of next generation multiple access (MA) schemes \cite{Tegos2020} has attracted considerable research interest.
Specifically, a promising way to break orthogonality is non-orthogonal multiple access where the power domain is utilized to provide MA.
In this case, multiuser detection techniques are required to retrieve the users' signals at the receiver, such as successive interference cancellation (SIC).
To this direction, rate-splitting multiple access (RSMA) provides flexible decoding and, thus, a more general and robust transmission framework in comparison with the conventional NOMA mechanism in \cite{Tegos2022}, which, in uplink, enables any point in the capacity region of the multiple access channel (MAC) to be achieved with successive decoding \cite{Rimoldi1996}.

Moreover, considering the characteristics of spectrum sharing in the modern MA framework,  the integration of MA and CR is expected to have a significant impact on coping with spectrum scarcity, as well as on meeting the satisfactory spectrum efficiency and the heterogeneity of QoS requirements.
Specifically, in \cite{Ding2020CL} and \cite{Ding2020CL2}, CR inspired NOMA was presented where a secondary user (SU) occupies the resource block of a primary user (PU) without impacting negatively the performance of the PU. However, some restricting assumptions regarding the decoding order and the power allocation were considered.
Furthermore, in \cite{Ding2021}, the outage probability error floor was partially avoided by adopting CR inspired SIC with hybrid decoding order which considers both CSI and QoS requirements. 
Moreover, a different hybrid SIC scheme inspired by CR was proposed in \cite{Sun2021} utilizing power control.
The concept of CR is also applied to RSMA in \cite{Liu2022_RSMA}.
However, in all existing works where the integration of MA and CR is studied, the main focus is on outage performance.
Notably,  although the main principle of CR, which is based on the provision of feedback to the SU regarding its transmit power, is compatible with the use of adaptive transmission rates, the latter has not been addressed in the existing literature. Specifically, the open problems include the optimal use of SIC with the aim to maximize the ergodic rate, i.e., the average highest achievable rate at which information can be transmitted with a negligible probability of error, and the analytical characterization of the corresponding performance.

Motivated by the above, in this work, two CR inspired protocols  based on RSMA and SIC for an uplink network, consisting of a BS and two users, i.e., a PU and an SU, are investigated in terms of ergodic rate.
Specifically, the considered protocols aim to serve the SU in a resource block which is originally allocated solely for the PU without negatively affecting the QoS of the PU. 
To this direction, we extract the ergodic rate of the SU considering a specific QoS for the PU for the two protocols, which is a metric that is more representative of the performance of the considered protocols compared with outage probability.
For the SIC protocol, we determine the decoding order that maximizes the ergodic rate of the SU and is different than the one proposed in \cite{Sun2021}.
Finally, simulations validate the theoretical analysis and also indicate the effectiveness of the considered protocols over conventional NOMA mechanisms, which are illustrated as benchmarks. 

\begin{figure*}[t!]
 \centering
  \subfigure[Transceiver framework based on RSMA.]{
    \label{Model_RSMA} 
    \includegraphics[width=0.48\linewidth]{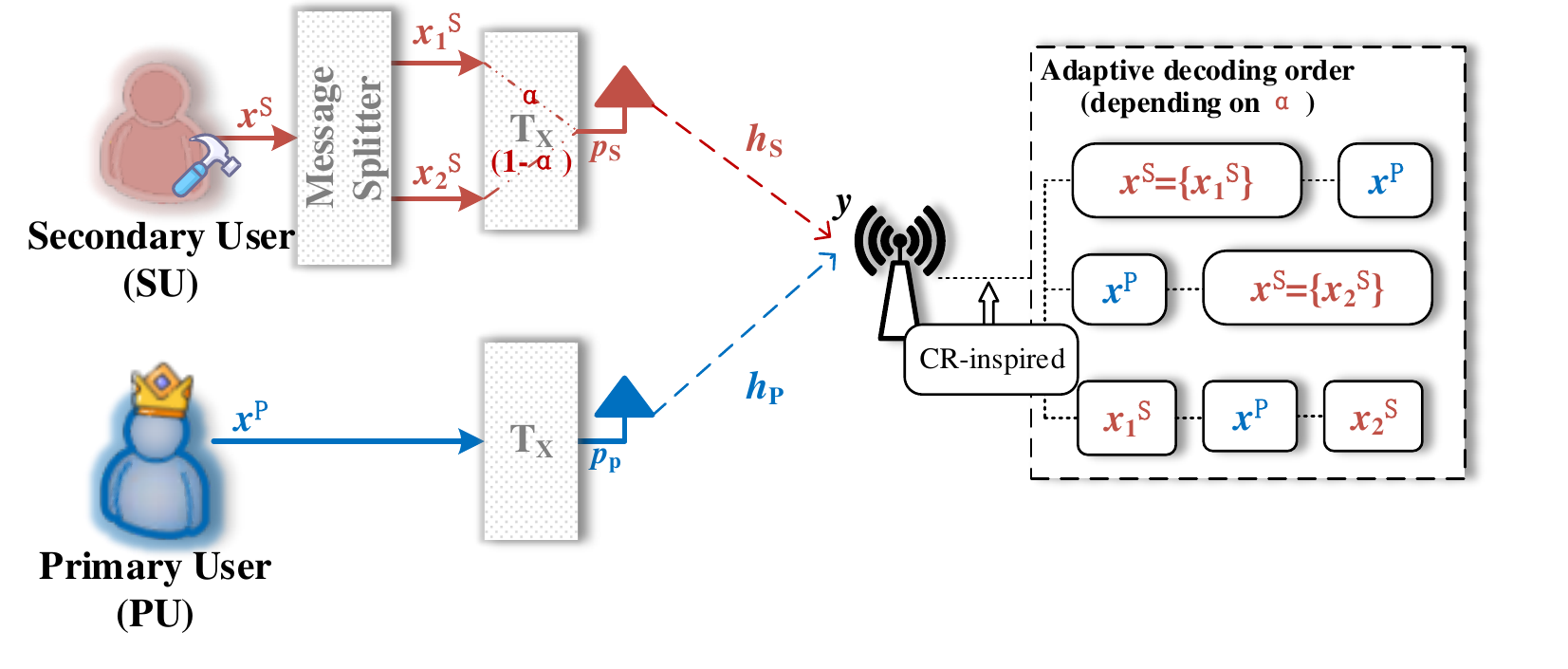}}
  \subfigure[Transceiver framework based on SIC.]{
    \label{Model_SIC} 
    \includegraphics[width=0.48\linewidth]{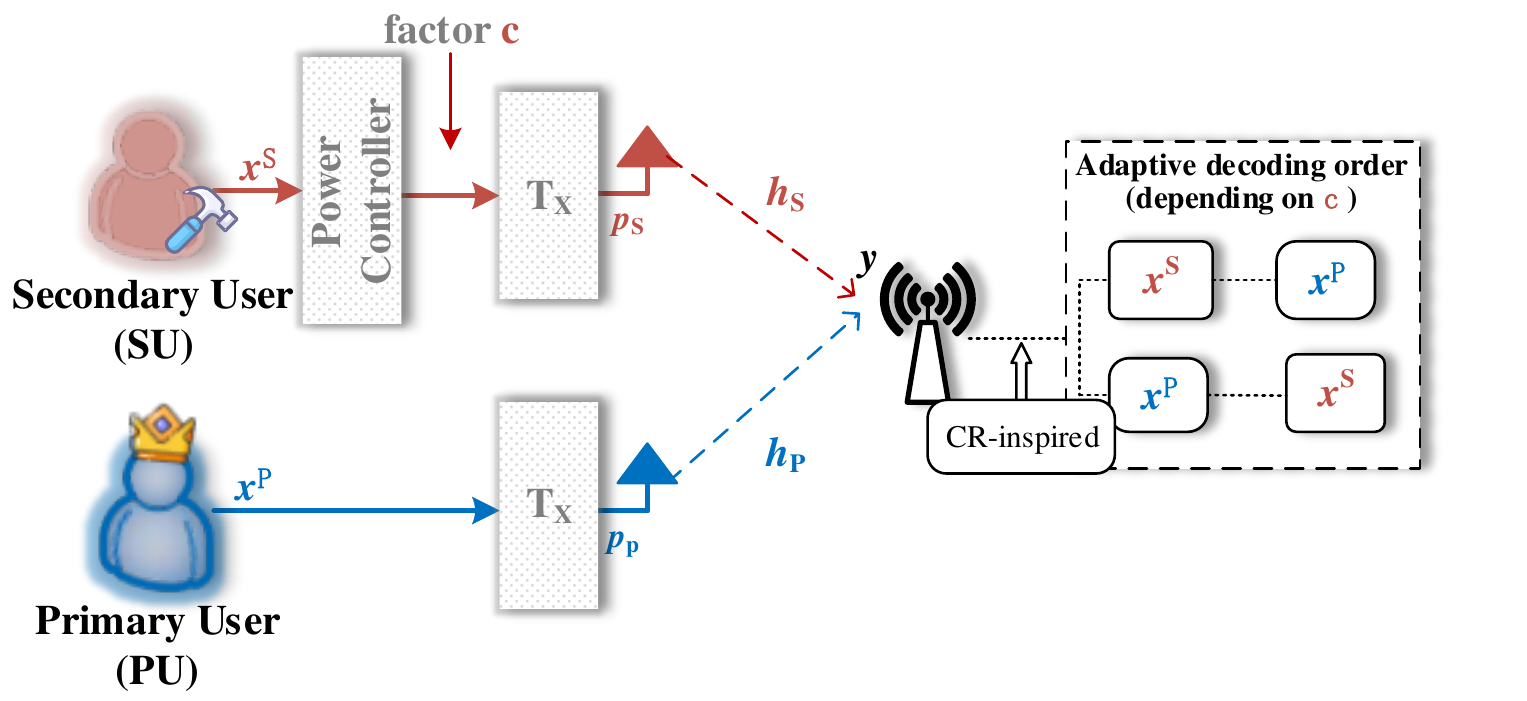}}
  \caption{Two CR inspired protocols.\vspace{-0.1in}}
  \label{fig:subfig} 
\end{figure*}


\section{System Model}

We consider an uplink network consisting of a BS and two users, a PU and an SU. All nodes are assumed to be equipped with a single antenna. For this network, two CR inspired protocols based on RSMA and SIC are leveraged, where the SU occupies the resource block of the PU without impacting negatively the performance of the PU.

The complex channel coefficient of the $i$-th user with $i\in\{\mathrm{P},\mathrm{S}\}$ is denoted by $h_i$. We assume Rayleigh fading, thus, $|h_i|^2$ follows the exponential distribution with rate parameter equal to one.
Furthermore, the received signal-to-noise ratio (SNR) at the BS from the $i$-th user can be denoted as 
\begin{equation}
{\gamma_i={\gamma}_{0i}l_i|h_i|^2},
\end{equation}
where ${\gamma}_{0i}$ and $l_i$ denote the average received SNR of the $i$-th user at the reference distance $d_0$ and the path loss coefficient between the $i$-th user and the BS, respectively. Specifically, $l_i$ is given by  
\begin{equation}
l_i=(\sfrac{d_i}{d_0})^{-u}
\end{equation}
with $d_i$ and $u$ being the distance between the $i$-th user and the BS and the path loss exponent, respectively. Therefore, the average received SNR $\gamma_{i}$ follows the exponential distribution with rate $\lambda_{i} = \frac{1}{{\gamma}_{0i}l_i}$.

\subsection{Cognitive radio inspired RSMA}
As  shown in Fig. \ref{Model_RSMA}, for the CR inspired RSMA protocol, i.e., CR-RSMA, the message of one user needs to be split in order to achieve any point of the capacity region \cite{Rimoldi1996}. 
It is assumed that the PU transmits a single message, while the SU transmits two messages in a way that does not have a negative impact to the performance of the PU.
Thus, the received message at the BS can be written as \cite{Liu2022_RSMA}
\begin{equation}\label{Rate_RSMA}
	y = \sqrt{\alpha l_\mathrm{S} p_\mathrm{S}} h_\mathrm{S} x_{1}^\mathrm{S} + \sqrt{(1-\alpha) l_\mathrm{S} p_\mathrm{S}} h_\mathrm{S} x_{2}^\mathrm{S}
     +\sqrt{l_\mathrm{P} p_\mathrm{P}} h_\mathrm{P} x^{\mathrm{P}} + n,
\end{equation}
where $p_i$ denotes the transmitted power of the $i$-th user. Moreover, $\alpha \in [0,1]$ and $n$ denote the power allocation factor and the additive white Gaussian noise at the BS with zero mean and variance $\sigma^2$, respectively. Furthermore, $x_{1}^\mathrm{S}$, $x_{2}^\mathrm{S}$ and $x^{\mathrm{P}}$ denote the first message of the SU, the second message of the SU and the message of the PU, respectively. Without loss of generality, it is assumed that $x_{1}^\mathrm{S}$ is decoded first. Thus, when decoding the message of the PU, $x_{2}^\mathrm{S}$ is handled as interference.
This decoding order provides flexibility and ensures that the performance of the PU is not negatively affected by properly selecting the power allocation factor $a$.

Therefore, the achievable rate of the PU can be written as
\begin{equation}\label{Rate_PU}
R_\mathrm{P} = B \log_2 \left(1+\frac{\gamma_\mathrm{P}}{(1-\alpha) \gamma_\mathrm{S} + 1} \right)
\end{equation}
with $B$ being the bandwidth of this system.
Since the performance of the PU must not be negatively affected, the constraint $\gamma_\mathrm{P} \geq \theta_\mathrm{P}$, where $\theta_\mathrm{i} = 2^{\sfrac{R_\mathrm{i,th}}{B}} - 1$ with ${R_\mathrm{i,th}}$ being the target rate, must be satisfied if possible. 
Thus, $\alpha$ is derived as follows
\begin{equation}\label{case_of_alpha}
	\alpha = \begin{cases}
		0, & \frac{\gamma_\mathrm{P}}{\gamma_\mathrm{S}+1} \geq \theta_\mathrm{P} \\
		1, & \gamma_\mathrm{P} < \theta_\mathrm{P} \\
		1 - \frac{1}{\gamma_\mathrm{S}} \left( \frac{\gamma_\mathrm{P}}{\theta_\mathrm{P}} - 1 \right), & \mathrm{otherwise.}
	\end{cases}
\end{equation}
The achievable rate of the SU is given by
\begin{equation}\label{SU_Oevents}
	\begin{aligned}
		R_\mathrm{S}&=B\log_2\left(1+\frac{\alpha \gamma_\mathrm{S}}{\gamma_\mathrm{P}+(1-\alpha)\gamma_\mathrm{S}+1}\right)\\
		&+B\log_2\left(1+(1-\alpha) \gamma_\mathrm{S}\right).
	\end{aligned}
\end{equation}
\subsection{Cognitive radio inspired SIC}
By using the main principles of CR, a similar protocol based on SIC, i.e., CR-SIC, is investigated which can be considered as a simpler special case of the above RSMA-based protocol illustrated in Fig. \ref{Model_SIC}.
In this SIC-based protocol, the SU transmits only one message utilizing a power allocation factor $c$ in order to avoid a negative impact to the performance of the PU.
Thus, the received message at the BS can be written as
\begin{equation}
	\begin{split}
		y = \sqrt{l_\mathrm{P} p_\mathrm{P}} h_\mathrm{P} x^{\mathrm{P}} + \sqrt{c l_\mathrm{S} p_\mathrm{S}} h_\mathrm{S} x^\mathrm{S} + n.
	\end{split}
\end{equation}
Similarly to CR-RSMA, the power allocation factor $c$ for the CR-SIC protocol in order to avoid a negative impact to the performance of the PU is given by
\begin{equation} \label{val_of_c2}
	c = \begin{cases}
		\frac{1}{\gamma_\mathrm{S}} \left( \frac{\gamma_\mathrm{P}}{\theta_\mathrm{P}} - 1 \right), & \theta_\mathrm{P} \leq \gamma_\mathrm{P} < \theta_\mathrm{P}(\gamma_\mathrm{S}+1) \\
		1, & \mathrm{otherwise.}
	\end{cases}
\end{equation}
It can be observed in \eqref{val_of_c2} that the power allocation factor of CR-SIC is equal to or less than one, thus, compared with CR-RSMA, CR-SIC can be considered as a more energy-efficient scheme.

The decoding order for CR-SIC, which is adopted in order to maximize the ergodic rate of the SU, considering a specific QoS requirement for the PU determined by  $\theta_\mathrm{P}$, is given by
\begin{equation} \label{DO2}
	\mathcal{F} = \begin{cases}
\{\mathrm{SU},\mathrm{PU}\}, & \gamma_\mathrm{P}\leq \theta_\mathrm{P} \ \text{or} \ \frac{\gamma_\mathrm{S}}{1+\gamma_\mathrm{P}}> \frac{\gamma_\mathrm{P}}{\theta_\mathrm{P}}-1, \gamma_\mathrm{P}> \theta_\mathrm{P} \\
	\{\mathrm{PU},\mathrm{SU}\}, & \text{otherwise}.
	\end{cases}
\end{equation}
It should be highlighted that, in \eqref{DO2}, the SU is decoded first, if the PU is in outage, or if the rate achieved by the SU by treating the message of the PU as interference is larger than the one achieved with $c<1$ as shown in \eqref{val_of_c2} when the SU is decoded second.
Also, \eqref{DO2} is different the one proposed in \cite{Sun2021} where the outage probability is minimized.

\section{Ergodic Rate}
In this section, the ergodic rate of the SU for the CR-RSMA and CR-SIC protocols are derived. To obtain more insights, the difference between these two CR inspired schemes is further discussed.

\subsection{Analysis of cognitive radio inspired RSMA}
Considering the power allocation factor given in (\ref{case_of_alpha}) and the achievable rate of the SU in (\ref{SU_Oevents}), the ergodic rate of SU in CR-RSMA can be expressed by
\begin{equation} \label{EC-RSMA}
\begin{aligned}
&\overline{C}_{\mathrm{R}}
=\mathbb{E}\left\{B\log_2\left(1+\frac{\gamma_\mathrm{S}}{1+\gamma_\mathrm{P}}\right)  \Big| \gamma_\mathrm{P}<\theta_\mathrm{P} \right\}\\
&+ \mathbb{E}\left\{B\log_2\left(\frac{1+\gamma_\mathrm{S}+\gamma_\mathrm{P}}{1+\theta_\mathrm{P}}\right) \Big| \theta_\mathrm{P} \leq\gamma_\mathrm{P}< \theta_\mathrm{P}\left(1+\gamma_\mathrm{S}\right)\right\}\\
&+\mathbb{E}\left\{B\log_2\left(1+ \gamma_\mathrm{S}\right) \big| \gamma_\mathrm{P}\geq \theta_\mathrm{P}\left(1+\gamma_\mathrm{S}\right) \right\}.
\end{aligned}
\end{equation}
\begin{theorem}\label{Theorem1}
The ergodic rate of the SU for the CR-RSMA protocol can be approximated by 
\begin{equation} \label{theor1}
	\begin{split}
	&\overline{C}_{\mathrm{R}} \approx
    \sum_{i=1}^n \mathcal{A}_i B\log_2 (1+\mu_i) \Phi_{z}(\mu_i) \\
    	&+\frac{B\lambda_{\mathrm{S}}(e^{\lambda_{\mathrm{P}}\theta_\mathrm{P}}-e^{-\lambda_\mathrm{P}\theta_\mathrm{P}})}{\ln2 (\lambda_{\mathrm{P}}\theta_\mathrm{P}+\lambda_{\mathrm{S}})}e^{\lambda_{\mathrm{P}}\theta_\mathrm{P}+\lambda_{\mathrm{S}}}E_i(-(\lambda_\mathrm{S}+\lambda_\mathrm{P}\theta_\mathrm{P}))\\
   & -\frac{Be^{\theta_\mathrm{P}(\lambda_{\mathrm{S}}-\lambda_{\mathrm{P}})+\lambda_{\mathrm{S}}}E_i(-\lambda_{\mathrm{S}}(\theta_\mathrm{P}+1))}{\ln2} +	\mathcal{C},
	\end{split}
\end{equation}
where $\mathcal{A}_i$ denotes the weights of function values at specified points in Gaussian quadrature. Specifically, $\mathcal{A}_i= w_i \mu_i^{-\alpha}e^{\mu_i}$, with $w_i$ and $\mu_i$ being the weights and nodes, respectively. Moreover, $\mu_i$ denotes the $i$-th root of Laguerre polynomial $L_n(\mu)$, $w_i$ is defined by $w_{i}={\frac {\mu_{i}}{\left(\left(n+1\right)L_{n+1}\left(\mu_{i}\right)\right)^{2}}}$\cite{Abramowitz1972}. $\mathrm{Ei}(\cdot)$ is the special function of exponential integral \cite{Gradshteyn2014}. In the first term of (\ref{theor2}), it stands 
\begin{equation} \label{EC-RSMA-J1_pdf}
\begin{aligned}
&\Phi_{z}(z)
=\frac{\lambda_\mathrm{S}\lambda_\mathrm{P}e^{-\lambda_\mathrm{S}z}}{\lambda_\mathrm{P}+\lambda_\mathrm{S}z} 
-\frac{\lambda_\mathrm{P}\lambda_\mathrm{S} e^{-\lambda_\mathrm{S}\theta_\mathrm{P}}e^{-\lambda_\mathrm{P}(1+\theta_\mathrm{P})z}}{(\lambda_\mathrm{P}+\lambda_\mathrm{S}z)^2}\\
&+\frac{\lambda_\mathrm{S}\lambda_\mathrm{P}e^{-\lambda_\mathrm{S}z}}{(\lambda_\mathrm{P}+\lambda_\mathrm{S}z)^2}
-\frac{\lambda_\mathrm{P}\lambda_\mathrm{S}(1+\theta_\mathrm{P})e^{-\lambda_\mathrm{P}\theta_\mathrm{P}}e^{-\lambda_\mathrm{S}(1+\theta_\mathrm{P})z}}{\lambda_\mathrm{P}+\lambda_\mathrm{S}z}
\end{aligned}
\end{equation}
and also 
\begin{equation}\label{C_item}
		\mathcal{C} = 
			\begin{cases}
				\begin{aligned}
					& \! \left(\frac{B \lambda_\mathrm{S} \theta_\mathrm{P}}{\ln 2} \!+\! \frac{\lambda_\mathrm{S} \theta_\mathrm{P} (\lambda_\mathrm{P} \theta_\mathrm{P}+1)}{\ln2 (1+\theta_\mathrm{P})\lambda_{\mathrm{P}}}\right)  e^{\lambda_{\mathrm{P}}}E_i\left(-\lambda_{\mathrm{P}}(\theta_\mathrm{P}+1)\right)\\
			&\qquad  +\frac{B\lambda_{\mathrm{P}}\theta_\mathrm{P}e^{-\lambda_{\mathrm{P}}\theta_\mathrm{P}}}{\ln2 (1+\theta_\mathrm{P})\lambda_{\mathrm{P}}},  \qquad \qquad \qquad \qquad   \lambda_{\mathrm{S}}=\lambda_{\mathrm{P}} \\	
			&\frac{B \lambda_\mathrm{S} e^{\lambda_\mathrm{S}}}{(\lambda_\mathrm{P}-\lambda_\mathrm{S})\ln 2}\Big(E_i\left(-(\lambda_\mathrm{S}+\lambda_\mathrm{P}\theta_\mathrm{P})\right)
			\\
			& \qquad -e^{\theta_\mathrm{P}(\lambda_\mathrm{P}-\lambda_\mathrm{S})}E_i\left(-(\lambda_\mathrm{S}+\lambda_\mathrm{S}\theta_\mathrm{P})\right)\Big), \quad \lambda_{\mathrm{S}}\neq\lambda_{\mathrm{P}}.
				\end{aligned}
			\end{cases}
 	\end{equation}
\end{theorem}
\begin{IEEEproof}
As shown in (\ref{EC-RSMA}), three terms are summed to derive the ergodic rate of the SU for CR-RSMA. 
The first term can be expressed as 
\begin{equation} \label{EC-SIC-J1}
\begin{aligned}
&\overline{C}_{\mathrm{R},J1}
\overset{z=\frac{x}{y+1}}=\int_0^\infty B\log_2 (1+z) \Phi_z(z) dz,
\end{aligned}
\end{equation}
where $\Phi_{z}(z)$ denotes the corresponding probability density function (PDF), which can be defined as
\begin{equation} 
\begin{aligned}
\Phi_{z}(z)=\frac{d [\Pr\{\gamma_\mathrm{S}<z(\gamma_\mathrm{P}+1), \gamma_\mathrm{P}<\theta_\mathrm{P}\} ]}{d z}.
\end{aligned}
\end{equation}
After algebraic manipulations, the closed-form expression of $\Phi_{z}(z)$ can be obtained as shown in (\ref{EC-RSMA-J1_pdf}). Next, Gauss-Laguerre quadrature can be leveraged to calculate the above function in terms of weighted summation of the dedicated function at specific nodes, thus the closed-form approximation is given by
\begin{equation} \label{EC-RSMA-J1_Gauss_Laguerre}
\begin{aligned}
&\overline{C}_{\mathrm{R},J1}=
\sum_{i=1}^n \mathcal{A}_i B\log_2 (1+\mu_i) \Phi_{z}(\mu_i).
\end{aligned}
\end{equation}
Next, the second item can be expressed as
\begin{equation} \label{EC-RSMA-J2}
\begin{aligned}
&\overline{C}_{\mathrm{R},J2} 
=\overset{z:=\gamma_\mathrm{P}+\gamma_\mathrm{S}}=\int_{\theta_\mathrm{P}}^{\infty}B\log_2 
 \left(\frac{1+z}{1+\theta_\mathrm{P}}
\right)\\
&\qquad \times \Bigg( \int_{0}^{\frac{z-\theta_\mathrm{P}}{1+\theta_\mathrm{P}}}\int_{\theta_\mathrm{P}}^{\theta_\mathrm{P}(x+1)} g(x,y)\ dy dx\\
&\qquad \qquad\qquad\qquad+\int_{\frac{z-\theta_\mathrm{P}}{1+\theta_\mathrm{P}}}^{z-\theta_\mathrm{P}}\int_{\theta_\mathrm{P}}^{z-x} g(x,y)\ dy dx \Bigg) dz,\\
\end{aligned}
\end{equation}
where $g(x,y)=\lambda_{\mathrm{S}}\lambda_{\mathrm{P}} e^{-\lambda_\mathrm{S} x} e^{-\lambda_\mathrm{P} y }$.
After some algebraic manipulations, (\ref{EC-RSMA-J2}) can be eventually expressed as
\begin{equation} \label{EC-RSMA-J2_CF}
\begin{aligned}
&\overline{C}_{\mathrm{R},J2}
= \frac{B\lambda_{\mathrm{S}}e^{\lambda_{\mathrm{P}}\theta_\mathrm{P}}}{\ln2 (\lambda_{\mathrm{P}}\theta_\mathrm{P}+\lambda_{\mathrm{S}})}e^{\lambda_{\mathrm{P}}\theta_\mathrm{P}+\lambda_{\mathrm{S}}}E_i(-(\lambda_\mathrm{S}+\lambda_\mathrm{P}\theta_\mathrm{P}))\\
&\qquad -\frac{Be^{\theta_\mathrm{P}(\lambda_{\mathrm{S}}-\lambda_{\mathrm{P}})+\lambda_{\mathrm{S}}}E_i(-\lambda_{\mathrm{S}}(\theta_\mathrm{P}+1))}{\ln2} +	\mathcal{C},
\end{aligned}
\end{equation}
where $\mathcal{C}$ is given by (\ref{C_item}).
Regarding the last term in (\ref{EC-RSMA}), there is 
\begin{equation} \label{EC-RSMA-J3}
\begin{aligned}
&\overline{C}_{\mathrm{R},J3}=B \int_o^\infty \int_{\theta_\mathrm{P}(x+1)}^\infty \log_2(1+x) g(x,y) dy dx\\
&\overset{(a)}=\frac{-B\lambda_{\mathrm{S}}e^{-\lambda_{\mathrm{P}}\theta_\mathrm{P}}}{\ln2 (\lambda_{\mathrm{P}}\theta_\mathrm{P}+\lambda_{\mathrm{S}})}e^{-\lambda_{\mathrm{P}}\theta_\mathrm{P}+\lambda_{\mathrm{S}}}E_i(-(\lambda_\mathrm{S}+\lambda_\mathrm{P}\theta_\mathrm{P})),
\end{aligned}
\end{equation}
where step (a) follows that  $\int_0^\infty\frac{e^{-\mu x}}{x+\beta}d_{x} = -e^{\mu \beta} \mathrm{Ei} (-\mu \beta)$  \cite{Gradshteyn2014}. To this end, the ergodic rate of the SU for CR-RSMA can be obtained as (\ref{theor1}),  by taking the summation of (\ref{EC-RSMA-J1_Gauss_Laguerre}), (\ref{EC-RSMA-J2_CF}), and (\ref{EC-RSMA-J3}), which completes the proof. 
\end{IEEEproof}

\subsection{Analysis of cognitive radio inspired SIC}
In this section, the ergodic rate of the SU for the CR-SIC scheme is derived considering a  decoding order aiming at the maximization of the ergodic rate of the SU, which is different than the one used for the outage probability in \cite{Sun2021}. 
Therefore, considering (\ref{DO2}) and (\ref{val_of_c2}), the ergodic rate of the SU for CR-SIC can be expressed as the union of four mutually exclusive events, i.e.,
\begin{equation} \label{EC-SIC-1}
\begin{split}
&\overline{C}_{\mathrm{S}}
=\mathbb{E}\left\{B\log_2\left(1+\frac{ \gamma_\mathrm{S}}{1+\gamma_\mathrm{P}}\right)\Big|\gamma_\mathrm{P}< \theta_\mathrm{P} \right\}
+ \mathbb{E}\bigg\{B \\
& \times \log_2\left(\frac{ \gamma_\mathrm{P}}{\theta_\mathrm{P}}\right) \Big| \theta_\mathrm{P} \leq\gamma_\mathrm{P}<\theta_\mathrm{P}(\gamma_\mathrm{S}+1), \frac{\gamma_\mathrm{S}}{1+\gamma_\mathrm{P}}\leq \frac{\gamma_\mathrm{P}}{\theta_\mathrm{P}}-1 \bigg\}\\
&+\mathbb{E}\left\{B\log_2\left(1+\frac{ \gamma_\mathrm{S}}{1+\gamma_\mathrm{P}}\right) \Big| \frac{\gamma_\mathrm{S}}{1+\gamma_\mathrm{P}}> \frac{\gamma_\mathrm{P}}{\theta_\mathrm{P}}-1, \gamma_\mathrm{P}> \theta_\mathrm{P} \right\}\\
&+ \mathbb{E}\left\{B\log_2\left(1+\gamma_\mathrm{S}\right)\big|\gamma_\mathrm{P}\geq\theta_\mathrm{P}(\gamma_\mathrm{S}+1) \right\}.
\end{split}
\end{equation}

\begin{theorem}\label{Theorem2}
The ergodic rate of the SU for the CR-SIC protocol can be approximated by 
\begin{equation} \label{theor2}
	\begin{split}
	\overline{C}_{\mathrm{S}} &\approx \sum_{i=1}^n \mathcal{A}_i B\log_2 (1+\mu_i) \Phi_{z}(\mu_i) + \sum_{i=1}^n \mathcal{B}_i \Xi(\mu_i)\\
	&+\sum_{i=1}^n{\sum_{j=1}^m}\mathcal{C}_i\mathcal{D}_j \Psi\big(\mu_i+\Theta(\mu_j+\theta_\mathrm{P}),~\mu_j+\theta_\mathrm{P}\big)\\
	&-\frac{B}{\ln 2}\frac{\lambda_\mathrm{S}e^{-\lambda_\mathrm{P}\theta_\mathrm{P}}}{\lambda_\mathrm{S}+\lambda_\mathrm{P}\theta_\mathrm{P}}e^{\lambda_\mathrm{S}+\lambda_\mathrm{P}\theta_\mathrm{P}} \mathrm{Ei}(-(\lambda_\mathrm{S}+\lambda_\mathrm{P}\theta_\mathrm{P})),
	\end{split}
\end{equation}
where $\mathcal{B}_i$, $\mathcal{C}_j$, and $\mathcal{D}_i$ denote the weights of function values at specified points in Gaussian quadrature, which are similar to $\mathcal{A}_i$ defined in Theorem. \ref{Theorem1}. 
In the second term of (\ref{theor2}), it stands
\begin{equation} \label{Xi_J3}
\begin{aligned}
&\Xi(x)=-\frac{\lambda_\mathrm{S}e^{-\lambda_\mathrm{S}x}}{\ln 2}\Big(\ln (x+1) e^{-\lambda_\mathrm{P}\theta_\mathrm{P}(x+1)}+\\
&\ln \frac{\theta_\mathrm{P}}{\tau(x)} e^{-\lambda_\mathrm{P}\tau(x)}+ \mathrm{Ei}\left(-\lambda_\mathrm{P}\tau\left(x\right)\right)-\mathrm{Ei}(-\lambda_\mathrm{P}\theta_\mathrm{P}(x+1))\Big),
\end{aligned}
\end{equation}
where $\tau(x)=\frac{\theta_\mathrm{P}-1+\sqrt{(\theta_\mathrm{P}+1)^2+4\theta_\mathrm{P}x}}{2}$, $\mathrm{Ei}(\cdot)$ is the special function of exponential integral \cite{Gradshteyn2014}.
In the third term,
$\Psi(x,y)=B\log_2 \left(1+\frac{x}{1+y}\right) g(x,y)$ with    $g(x,y)=\lambda_{\mathrm{S}}\lambda_{\mathrm{P}} e^{-\lambda_\mathrm{S} x} e^{-\lambda_\mathrm{P} y }$, and $\Theta(y)=(1+y)\left(\frac{y}{\theta_\mathrm{P}}-1\right)$.
\end{theorem}
\begin{IEEEproof}
As shown in (\ref{EC-SIC-1}), four terms are summed to derive the ergodic rate of the SU for the CR-SIC protocol.
Specifically, the first term is the same as the that of $\overline{C}_{\mathrm{R},J1}$.
Subsequently, the second term can be expressed as
\begin{equation} \label{EC-SIC-J2}
\begin{aligned}
&\overline{C}_{\mathrm{S},J2}
=\int_{0}^{\infty}\underbrace{\int_{\tau(x)}^{\theta_\mathrm{P}(x+1)}B\log_2 
 \left(\frac{y}{\theta_\mathrm{P}}   
\right)  g(x,y)\ dy}_{{\Xi(x)}}  dx,
\end{aligned}
\end{equation}
where the integral $\Xi(x)$ is derived in closed-form and is given by (\ref{Xi_J3}).
Similarly, (\ref{EC-SIC-J2}) can be further expressed as 
\begin{equation} \label{EC-SIC-J2_app}
\begin{aligned}
\overline{C}_{\mathrm{S},J2}
&\approx \sum_{i=1}^n \mathcal{B}_i \Xi(\mu_i).
\end{aligned}
\end{equation}
Next, considering the pre-conditions in (\ref{EC-SIC-1}), the third term is given by 
\begin{equation} \label{EC-SIC-J3}
\begin{aligned}
\overline{C}_{\mathrm{S},J3} =\int_{\theta_\mathrm{P}}^{\infty}\int_{\Theta(y)}^{\infty} \Psi(x,y)\ dx  dy,
\end{aligned}
\end{equation}
where $\Psi(x,y)=B\log_2 \left(1+\frac{x}{1+y}\right) g(x,y)$. Then, taking into account the modified Gauss-Laguerre quadrature, (\ref{EC-SIC-J2}) can be approximated as
\begin{equation} \label{EC-SIC-J3_app_double}
\begin{aligned}
\overline{C}_{\mathrm{S},J3}
&\approx\sum_{i=1}^n{\sum_{j=1}^m}\mathcal{C}_i\mathcal{D}_j \Psi\big(\mu_i+\Theta(\mu_j+\theta_\mathrm{P}),~\mu_j+\theta_\mathrm{P}\big).
\end{aligned}
\end{equation}
Eventually, it is noted that the last event in (\ref{EC-SIC-1}) is the same as $\overline{C}_{\mathrm{R},J3}$, which completes the proof.
\end{IEEEproof}
\subsection{Comparison of CR-RSMA and CR-SIC schemes}
Based on the above analysis for the investigated CR inspired schemes in terms of the ergodic rate, the difference between CR-RSMA and CR-SIC solely depends on the second term of (\ref{EC-RSMA}), i.e., $\overline{C}_{\mathrm{R},J2}$, and the second as well as the third term of (\ref{EC-SIC-1}), i.e., $\overline{C}_{\mathrm{S},J2}$ and $\overline{C}_{\mathrm{S},J3}$. Thus, the difference in terms of the ergodic rate between $\overline{C}_{\mathrm{R}}$ and $\overline{C}_{\mathrm{S}}$ can be evaluated as 
\begin{equation} \label{Difference_EC}
\begin{aligned}
\Delta_{\overline{C}}&=\overline{C}_{\mathrm{R}}-\overline{C}_{\mathrm{S}}\\
&=\mathbb{E}\left\{B\log_2\left(\frac{1+\gamma_\mathrm{S}+\gamma_\mathrm{P}}{1+\theta_\mathrm{P}}\right) \Big| \theta_\mathrm{P} \leq\gamma_\mathrm{P}< \theta_\mathrm{P}\left(1+\gamma_\mathrm{S}\right)\right\}\\
&-\mathbb{E}\bigg\{B \log_2\left(\frac{ \gamma_\mathrm{P}}{\theta_\mathrm{P}}\right) \Big| \Omega({\theta_\mathrm{P}},\gamma_\mathrm{S})<\gamma_\mathrm{P}< \theta_\mathrm{P}\left(1+\gamma_\mathrm{S}\right) \bigg\}\\
&-\mathbb{E}\left\{B\log_2\left(1+\frac{ \gamma_\mathrm{S}}{1+\gamma_\mathrm{P}}\right) \Big| \theta_\mathrm{P} \leq\gamma_\mathrm{P}\leq \Omega({\theta_\mathrm{P}},\gamma_\mathrm{S}) \right\},
\end{aligned}
\end{equation}
where $\Omega({\theta_\mathrm{P}},\gamma_\mathrm{S})=\frac{\theta_\mathrm{P}-1+\sqrt{(\theta_\mathrm{P}-1)^2+4\theta_\mathrm{P}(1+\gamma_\mathrm{S})}}{2}$.
\begin{proposition}\label{Prop_1}
If $\gamma_\mathrm{P}\in[\theta_\mathrm{P}, \theta_\mathrm{P}\left(1+\gamma_\mathrm{S}\right)]$, CR-RSMA can achieve larger ergodic rate than the CR-SIC scheme. In other words, when $c=\frac{1}{\gamma_\mathrm{S}} \left( \frac{\gamma_\mathrm{P}}{\theta_\mathrm{P}} - 1 \right)<1$ as shown in (\ref{val_of_c2}), it always holds that $\overline{C}_{\mathrm{R}}>\overline{C}_{\mathrm{S}}$, otherwise CR-RSMA and CR-SIC can obtain the same ergodic rate. 
However, when $c<1$ in CR-SIC, less power is transmitted compared with CR-RSMA.
\end{proposition}
\begin{IEEEproof}
By splitting the range of $\gamma_\mathrm{P}\in[\theta_\mathrm{P}, \theta_\mathrm{P}\left(1+\gamma_\mathrm{S}\right)]$ into $\gamma_\mathrm{P}\in[\theta_\mathrm{P}, \Omega(\cdot)]$ and $\gamma_\mathrm{P}\in[\Omega(\cdot), \theta_\mathrm{P}\left(1+\gamma_\mathrm{S}\right)]$, it stands that $\frac{1+\gamma_\mathrm{S}+\gamma_\mathrm{P}}{1+\theta_\mathrm{P}}>\frac{1+\gamma_\mathrm{S}+\gamma_\mathrm{P}}{1+\gamma_\mathrm{P}}$ and $\frac{1+\gamma_\mathrm{S}+\gamma_\mathrm{P}}{1+\theta_\mathrm{P}}>\frac{ \gamma_\mathrm{P}}{\theta_\mathrm{P}}$, respectively, thus resulting in $\Delta_{\overline{C}}>0$, which completes the proof.
\end{IEEEproof}

\section{Simulation Results}
\begin{figure} 
\setlength{\abovedisplayskip}{0pt}
\setlength{\belowdisplayskip}{0pt}
\centering
\includegraphics[width=0.9\linewidth]{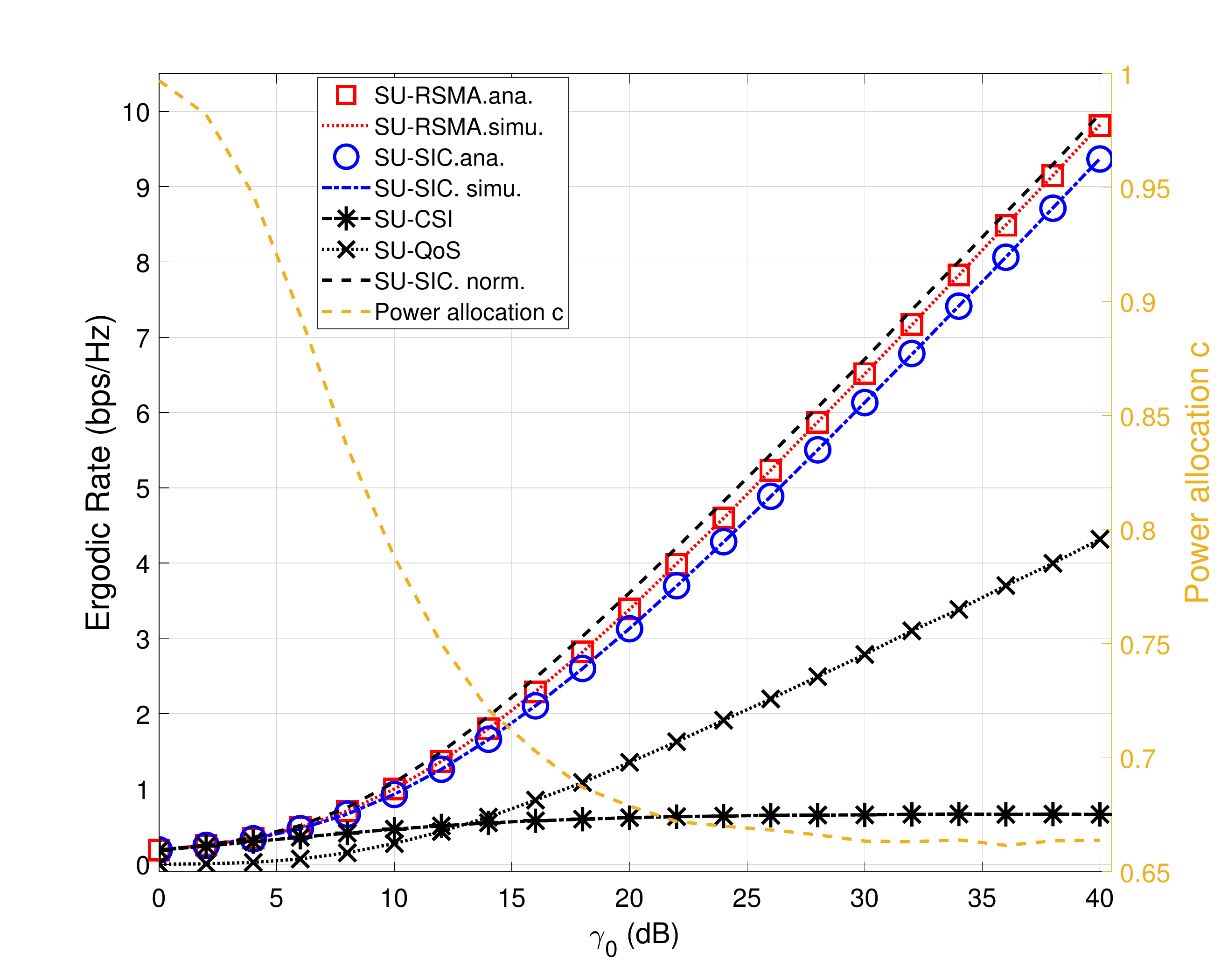}
\caption{Ergodic rate versus $\gamma_\textrm{0}$. }
\label{[OP_rho_SU_PU]}
\end{figure}
In this section, Monte Carlo simulations are performed to validate the derived expressions of the corresponding ergodic rate of CR-RSMA and CR-SIC. Moreover, the results also reveal the effectiveness of the investigated protocols compared with the conventional NOMA-based methods in \cite{Ding2020CL}, i.e., a two-user uplink NOMA system with CSI-based SIC and QoS-based SIC, termed ``CSI"  and ``QoS", respectively. 
Specifically, ``CSI" scheme denotes the case that the SU is permitted to access the channel, via which the PU transmits as if it solely occupied, i.e., $\overline{C}_C=\mathbb{E}\left\{B\log_2\left(1+\frac{ \gamma_\mathrm{S}}{1+\gamma_\mathrm{P}}\right)\right\}$.
In the ``QoS" scheme, the PU is decoded first and the SU is treated as interference, thus $\overline{C}_Q=\mathbb{E}\left\{B\log_2\left(1+\gamma_\mathrm{S}\right)\left|\gamma_\mathrm{P}>\theta_\mathrm{P}(\gamma_\mathrm{S}+1)\right.\right\}$. 
In what follows, unless stated otherwise, we set the distances of the two users $\sfrac{d_{\rm{P}}}{d_0}=1$ and $\sfrac{d_{\rm{S}}}{d_0}=2$, the path loss exponent $u=2$, and the normalized target rates $\sfrac{R_{\mathrm{P, th}}}{B}=2.5$ bps/Hz.

As illustrated in Fig. \ref{[OP_rho_SU_PU]}, the ergodic rate of the SU is plotted versus different SNR values $\gamma_{\textrm{0S}}=\gamma_{\textrm{0P}} = \gamma_\textrm{0}$ for CR-RSMA, CR-SIC, and the two baselines.
It can be observed that the simulations and the analytical results coincide, and that CR-RSMA and CR-SIC outperform the baselines. 
Meanwhile, with the selected parameters CR-RSMA always outperforms CR-SIC in terms of the ergodic rate, while $c$ decreases with the increase of $\gamma_0$, as presented in Proposition \ref{Prop_1}. 
Moreover, since, in the CR-SIC scheme,  less power may be transmitted compared with CR-RSMA, as illustrated by the line of $c$, a modified CR-SIC, where the transmitted SNR is normalized with respect to $c$, is also plotted, namely ``SU-SIC, norm". It can be observed that the ``SU-SIC, norm" can obtain larger ergodic rate than CR-RSMA, where the power allocation factors are summed to 1, thus no normalization is necessary.


\begin{figure} 
\setlength{\abovedisplayskip}{0pt}
\setlength{\belowdisplayskip}{0pt}
\centering
\includegraphics[width=0.9\linewidth]{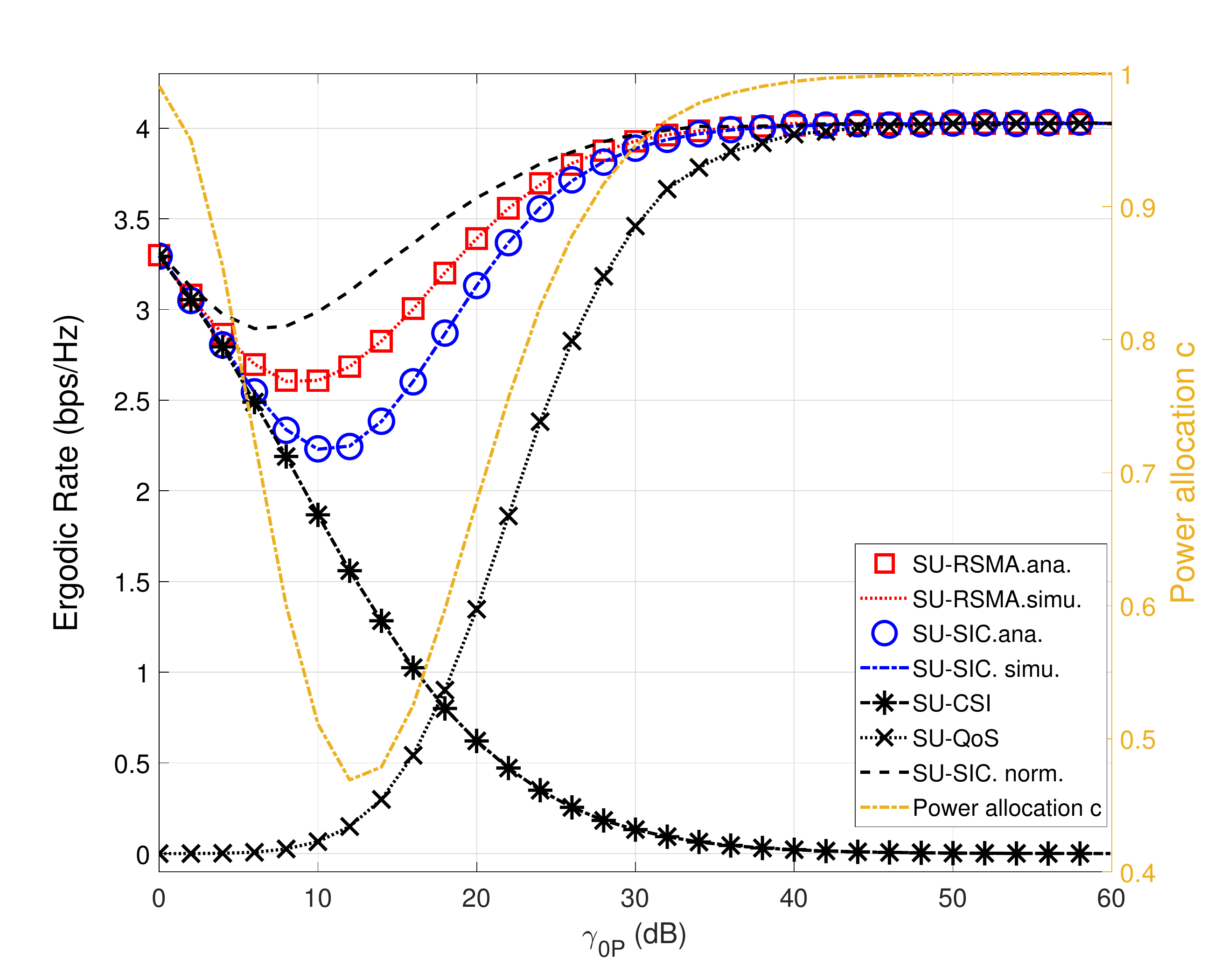}
\caption{Ergodic rate versus  $\gamma_\textrm{0P}$ with $\gamma_\textrm{0S}=20$ dB. }
\label{EC_rho_SU_PU}
\end{figure}

To obtain more insights on the CR-RSMA and CR-SIC protocols, the ergodic rate versus the transmitted SNR $\gamma_{\rm{0P}}$ with  $\gamma_{\rm{0S}}=20$ dB is depicted in Fig. \ref{EC_rho_SU_PU}. Following the definition of ``CSI" and ``QoS" benchmarks, they vary inversely with the increasing $\gamma_{\rm{0P}}$, i.e., the former scheme decreases due to the constant value of $\gamma_{\rm{0S}}$, while the latter one increases to achieve   $B\log_2\left(1+\gamma_\mathrm{S}\right)$ by also meeting the condition of $\gamma_\mathrm{P} > \theta_\mathrm{P}(\gamma_\mathrm{S}+1)$. Regarding to the CR-RSMA and CR-SIC protocols, both of them first decrease, then increase,  and eventually achieve the same result as ``QoS”, i.e., $B\log_2\left(1+\gamma_\mathrm{S}\right)$, because of the condition $\gamma_\mathrm{P}\geq\theta_\mathrm{P}(\gamma_\mathrm{S}+1)$ as $\gamma_\mathrm{P}$ increases. However, the superiority of the considered schemes in low and medium SNR compared with the ``QoS'' scheme is evident. Moreover, when $c<1$, CR-RSMA outperforms CR-SIC, while they illustrate the same performance when $c=1$, which also  validates Proposition \ref{Prop_1}. Eventually, the normalized CR-SIC is also considered to obtain a fair comparison with CR-RSMA.

\section{Conclusions} \label{[Conclusion]}
In this work, an uplink network consisting of one BS, a PU, and an SU, has been considered and two CR inspired protocol based on RSMA and SIC have been investigated. 
Without impacting negatively the performance of the PU, the SU shares the communication resource block which is originally solely occupied by the PU.
To this point, we have derived the ergodic rate of the SU for the two considered protocols by utilizing the optimal decoding order.
Finally, numerical results have been presented to demonstrate the accuracy of the theoretical analysis and illustrate the superiority of the investigated  protocols over two benchmark NOMA schemes also inspired by CR.

\bibliographystyle{IEEEtran}
\bibliography{Bibliography}


\end{document}